\newcommand{\CLASSINPUTtoptextmargin}{2.4cm}
\newcommand{\CLASSINPUTbottomtextmargin}{4.6cm}
\documentclass[conference]{IEEEtran}
\IEEEoverridecommandlockouts

\usepackage{authblk}
\usepackage{cite}
\usepackage{amsmath,amssymb,amsfonts}
\usepackage{amsmath}  % for mathematical symbols like \gets
\usepackage{algorithm}  % for algorithm environment
\usepackage{algorithmicx}  % for algorithmic environment
\usepackage{algpseudocode}  % for pseudocode formatting
\usepackage{amsfonts}  % for special mathematical fonts (if needed)
\usepackage{graphicx}
\usepackage{textcomp}
\usepackage[table]{xcolor}

\usepackage{tabularx}
\usepackage{url}
\usepackage[bookmarks=false]{hyperref}
\usepackage{rotating}
\usepackage{enumitem}
\usepackage{subcaption}
\usepackage{multirow}
\usepackage{pifont}
\usepackage{tcolorbox}
\definecolor{tumColorLightBlue}{HTML}{f0f5fa} 
\usepackage{cleveref}
\usepackage{makecell}

\definecolor{designBlue}{RGB}{91,155,213}
\definecolor{configurationGreen}{RGB}{112,173,71}
\definecolor{validationPurple}{RGB}{165,143,190}
\definecolor{deploymentYellow}{RGB}{255,192,0}
\definecolor{operationRed}{RGB}{237,125,122}

\usepackage{tablefootnote}
\usepackage{threeparttable}
\usepackage{multirow}
\usepackage{booktabs}
\usepackage{harveyballs}

\usepackage{tcolorbox}
\definecolor{tumColorLightBlue}{HTML}{f0f5fa} 
\usepackage{cleveref}

\IEEEspecialpapernotice{Special Session Paper}

\renewcommand{\baselinestretch}{0.94}

\usepackage{eso-pic}
\usepackage[outdir=./]{epstopdf}
\usepackage{fancyhdr}

\def\BibTeX{{\rm B\kern-.05em{\sc i\kern-.025em b}\kern-.08em
    T\kern-.1667em\lower.7ex\hbox{E}\kern-.125emX}}

\AddToShipoutPictureBG*{%
  \AtPageUpperLeft{%
    \put(0.07\paperwidth,-1.2cm){%
      \fbox{%
        \begin{minipage}{\textwidth}
          \centering
          \small
          \textbf{This paper has been accepted for publication in the proceedings of the 2026 IEEE/ACM International Conference on Embedded Software (EMSOFT), October 2026. The copyright for this paper will be transferred to IEEE. ©2026 IEEE. Personal use of this material is permitted. Permission from IEEE must be obtained for all other uses, in any current or future media, including reprinting/republishing this material for advertising or promotional purposes, creating new collective works, for resale or redistribution to servers or lists, or reuse of any copyrighted component of this work in other works.}
        \end{minipage}
      }%
    }%
  }%
}

\title{Making Time-Sensitive Networking Deployable: \\ A Comprehensive Lifecycle Architecture}

\author[1]{Rubi Debnath}
\author[2]{Paul Pop}
\author[2,3]{Silviu S. Craciunas}
\author[4]{Marc Boyer}
\author[1]{Sebastian Steinhorst} 

\affil[1]{TUM School of Computation, Information and Technology, Technical University of Munich, Germany}
\affil[2]{DTU Compute, Technical University of Denmark, Denmark}
\affil[3]{NXP Semiconductors, Vienna, Austria}
\affil[4]{DTIS, ONERA, Universit\'e de Toulouse, 31000, Toulouse, France}
{
    \makeatletter
    \renewcommand\AB@affilsepx{, \protect\Affilfont}
    \makeatother
    \affil[1]{firstname.lastname@tum.de}
    \affil[2]{paupo@dtu.dk}
    \affil[3]{silviu.craciunas@nxp.com}
    \affil[4]{marc.boyer@onera.fr}
}
\makeatletter
\patchcmd{\@maketitle}
  {\addvspace{0.5\baselineskip}\egroup}
  {\addvspace{-0.5\baselineskip}\egroup}
  {}
  {}

\def\bstctlcite{\@ifnextchar[{\@bstctlcite}{\@bstctlcite[@auxout]}}
\def\@bstctlcite[#1]#2{\@bsphack
  \@for\@citeb:=#2\do{%
    \edef\@citeb{\expandafter\@firstofone\@citeb}%
    \if@filesw
      \immediate\write\csname #1\endcsname{\string\citation{\@citeb}}%
    \fi}%
  \@esphack}
  
\makeatother

\begin{document}
\maketitle

\begin{abstract}
Time-Sensitive Networking (TSN) provides deterministic guarantees and bounded latency for safety-critical applications across different domains. While the IEEE 802.1 TSN standards provide a comprehensive set of mechanisms for time synchronization, shaping, scheduling, resource reservation, and reliability, deploying TSN in real-world applications remains a complex, multi-step engineering task. Existing research primarily focuses on optimizing individual objectives, such as increasing schedulability, reducing latency, accelerating solution generation, or improving performance analysis. However, whether these extensively researched solutions can be deployed in real-world systems remains unanswered, as the assumptions made in these methodologies often differ from practical hardware constraints. Therefore, a broader, holistic view of end-to-end TSN deployment and management remains an open research question. Currently, there is a lack of a unified perspective, consensus, and coordinated research efforts focusing on a comprehensive end-to-end TSN deployment workflow. We present a comprehensive overview of the TSN deployment lifecycle, current challenges, limitations of existing tools, and future research directions for TSN deployment and management. Lastly, we identify key research gaps from a deployment perspective and provide guidance for the development of next-generation deployable TSN networks.
\end{abstract}

\section{Introduction}
\label{sec:introduction}
Time-Sensitive Networking (TSN) is a set of IEEE 802.1 sub-standards that provides deterministic communication capabilities, including time synchronization, bounded latency and jitter, resource reservation, and reliability mechanisms. TSN has a wide range of applications, including automotive systems~\cite{tsn_automotive}, industrial automation~\cite{tsn_industrial_automation}, spacecraft~\cite{tsn_space}, and healthcare~\cite{tsn_telemedicine}. To support different application requirements, TSN provides multiple scheduling and shaping mechanisms, including Time-Aware Shaper (TAS) standardized by IEEE 802.1Qbv~\cite{8021Qbv}, Cyclic Queuing and Forwarding (CQF)~\cite{8021Qch}, Cycle Specific Queuing and Forwarding (CSQF)~\cite{ietfSegmentRouting}, Multi-CQF~\cite{norman_2019,mcqf_paul,rubi_iotj}, Credit-Based Shaper (CBS)~\cite{8021QAV}, Asynchronous Traffic Shaper (ATS)~\cite{8021Qcr,specht_ubs}, and Frame Preemption (FP)~\cite{8021Qbu,rubi_noms}. Among these, TAS has been extensively studied because it enables deterministic transmission of Time-Triggered (TT) flows, also referred to as Scheduled Traffic (ST), using a Gate Control List (GCL). However, the generation of GCLs is an NP-hard problem~\cite{CraciunasRTNS16}, and the complexity increases with the number of flows and network size~\cite{ilp_limitations}.

\begin{figure}[!t]
    \centering
    \includegraphics[width=\linewidth, trim={0cm 0cm 0cm 0cm}, clip]{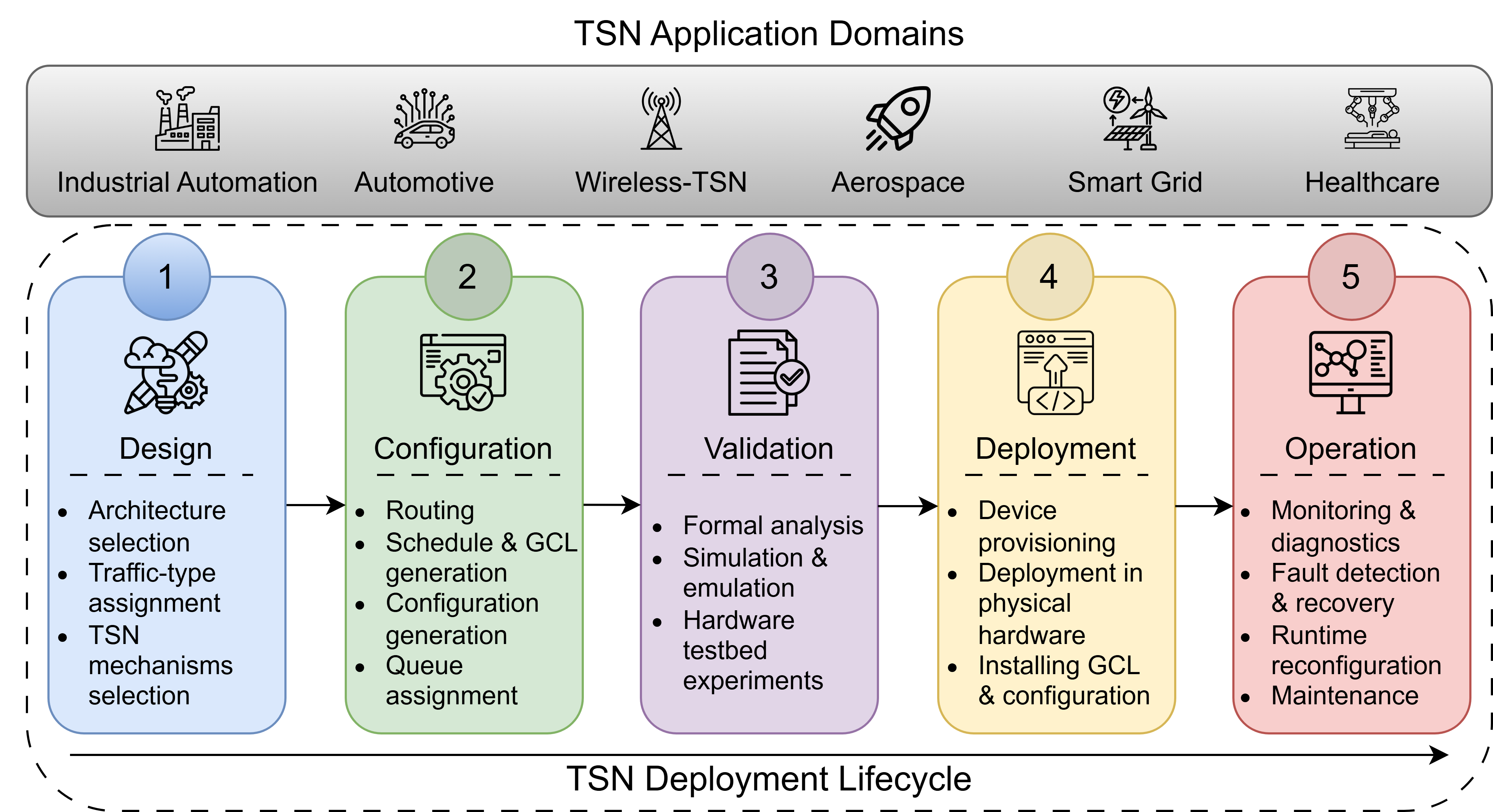}
    \caption{Overview of the TSN deployment lifecycle discussed in this paper. The lifecycle consists of five sequential stages: design, configuration, validation, deployment, and operation.}
    \label{fig:fig1}
\end{figure}

Although significant progress has been made in TSN schedulability analysis, configuration, and optimization, deploying TSN in real-world systems remains a complex multi-step engineering challenge. Existing research primarily focuses on individual optimization problems, e.g., improving schedulability~\cite{rubi_iotj}, reducing latency~\cite{mcqf_paul}, accelerating configuration synthesis, or performing worst-case delay analysis~\cite{luxi_avb_rtas, luxi_multiple_avb}. These approaches have significantly advanced the theoretical foundations of TSN. However, whether solutions developed in research can be directly integrated into real-world deployments remains an open question. The assumptions made by optimization and analysis methods often differ from practical deployment constraints, including hardware limitations, configuration interfaces, device capabilities, and operational requirements.

Practical TSN deployment is not limited to solving a scheduling problem or finding a suitable configuration. A system engineer must determine the appropriate TSN architecture, select suitable combinations of mechanisms, assign traffic types, synthesize schedules, configure network devices, validate the resulting configuration, deploy it to physical infrastructure, and continuously manage the network during operation. In addition, TSN networks support mixed-shaper architectures, where multiple mechanisms such as TAS, CBS, ATS, and CQF coexist at the same egress port. While such combinations can provide improved flexibility, they also introduce additional interactions between mechanisms, increasing the complexity of configuration and verification~\cite{luxi_tnsm,paul_iet_multi_shaper,rubi_rtcsa,rubi_noms,voica_multi_shaper_tsn}. Moreover, selecting the appropriate combination of mechanisms from a pool of different shapers and schedulers for a given input becomes a crucial design choice.

Therefore, TSN deployment should be considered as an end-to-end lifecycle rather than as a collection of isolated optimization problems (refer to Fig.~\ref{fig:fig1}). Each stage represents a set of decisions required to transform application requirements into an operational TSN network. The output of each stage becomes the input for the subsequent stage. Therefore, the decisions made during early stages directly influence later validation, deployment, and runtime management. 

To bridge the gap between TSN research and practical deployment, this paper presents a comprehensive lifecycle-oriented analysis of TSN deployment and management. The main objective of this paper is to establish a unified perspective on deployable TSN networks and highlight the research directions required to move from TSN mechanism optimization toward complete end-to-end TSN deployment workflows. 

This paper starts by discussing the background of the TSN deployment lifecycle (refer to Section~\ref{sec:background}), highlights the challenges of TSN deployment and runtime operation (refer to Section~\ref{sec:deployment_and_runtime_challenges}), discusses the gaps and limitations of current TSN profiles and standardization efforts (refer to Section~\ref{sec:configuration_and_standardization_efforts}), and presents runtime reconfiguration challenges and deployment-oriented solutions (refer to Section~\ref{sec:reconfiguration_by_marc_boyer}). Finally, we discuss automation-aided TSN deployment and the role of Artificial Intelligence (AI) and Machine Learning (ML) in TSN management and configuration (refer to Section~\ref{sec:llm_tsn_deployment}). 

\section{TSN Deployment Lifecycle: From Design to Operation}
\label{sec:background}
The TSN deployment lifecycle consists of five sequential stages, in which the output of one stage is used as the input to the subsequent stage. Fig.~\ref{fig:fig1} presents the different stages involved in end-to-end TSN management and deployment. We broadly categorize these stages into five phases: (1) design, (2) configuration, (3) validation, (4) deployment, and (5) operation. Each of these stages is highly complex and involves ongoing research on optimization. Therefore, the overall configuration space is large. Moreover, the configuration process is still poorly understood by the community. System engineers must navigate numerous interdependent decisions while accounting for the hardware limitations of commercial switches and satisfying end-to-end timing requirements. Multi-domain and wireless TSN networks, as well as runtime reconfiguration, further increase this complexity. Current tools are sparsely developed, are rarely open sourced, and seldom scale. Finding a feasible configuration for medium-scale networks requires person-months of engineering effort. In addition, precise time synchronization remains a fundamental requirement for  many TSN mechanisms. Therefore, large-scale industrial deployment requires addressing several challenges across the TSN deployment lifecycle~\cite{tsn_industrial_automation}. We discuss these stages below. 

\subsection{Design Phase}
In the design stage, the requirements and architecture of the TSN network are defined before configuring the network devices. Here, network designers must determine the appropriate combination of TSN mechanisms and perform traffic-type assignment (TTA) based on the quality-of-service (QoS) requirements. There is no universal solution applicable to all domains. Therefore, different applications, such as industrial automation, automotive, and aerospace systems, require customized solutions based on their specific requirements. Previous work has used metaheuristics~\cite{voica_traffic_assignment} and Deep Reinforcement Learning (DRL)-based~\cite{rubi_icc} solutions to solve the isolated TTA problem. 

\subsection{Configuration Phase}
The configuration stage is one of the most challenging, as the standard leaves the configuration process to the implementation. This stage requires configuring several parameters for a simple TSN network, including time synchronization, schedule generation, queue assignment, cycle duration selection, bandwidth (BW) reservation, redundancy, and/or resource allocation. For TAS, optimized algorithms have been developed to generate the GCL~\cite{silviu_gate_array, window_niklas, CraciunasRTNS16}. For CQF, CSQF, and Multi-CQF, cycle duration selection, BW distribution, and configuration generation become crucial~\cite{boyer_cycle_duration, deepscheduler, deepcqf_icc_cheng, itp, rubi_iotj, mcqf_paul}. 

\subsection{Validation Phase}
Before deploying the network in real-world scenarios, thorough validation of the design and configuration is necessary to avoid network downtime. The validation stage ensures that the generated TSN configuration satisfies the QoS requirements before deployment in an operational environment. Existing research employs multiple validation approaches, including analytical methods, simulation frameworks~\cite{rubi_vtc}, formal verification techniques, and physical testbeds~\cite{tsn_flextest_frank}. Analytical methods provide Worst-Case Delay (WCD) estimation and schedulability analysis of the flows in the network using tools such as Network Calculus (NC)~\cite{luxi_multiple_avb, luxi_tnsm}. Formal verification techniques, such as SMT and ILP, provide WCD estimates. Nevertheless, NC-based approaches are often highly pessimistic, and formal mathematical methods may not fully capture real-world scenarios. Therefore, simulation-based validation~\cite{rubi_rtcsa, rubi_noms, rubi_ccnc, rubi_vtc} is used to complement NC and formal verification approaches. Since simulation results may not fully represent hardware implementation scenarios~\cite{tsn_flextest_frank}, hardware testbeds remain important for validation.

\subsection{Deployment Phase}
In the deployment phase, the validated configuration is transferred to the physical network infrastructure. A desirable future TSN capability is an automated one-touch or one-click deployment solution for the entire network, avoiding device-by-device configuration and simplifying large-scale cross-domain deployments. Automated support for debugging and configuring network devices is also highly desirable. 

\subsection{Operation Phase}
The operation stage consists of the continuous management and monitoring of an operational TSN network. The goal is to monitor network logs, detect malicious activities, adapt the network according to new flows or faulty nodes, and/or detect timing violations. Another major need is the dynamic reconfiguration during runtime for networks under operation.

\section{Why TSN Remains Difficult to Deploy}
\label{sec:deployment_and_runtime_challenges}
TSN deployment is more than a technical problem. In multi-domain systems, domains may be operated by different teams, departments, or companies, creating a systems-of-systems challenge that includes governance, ownership, and rules defining who may change each configuration and when. Although these concerns affect deployability, this paper focuses on the technical challenges. Industrial applications carry flows with widely varying timing requirements. A control loop may require short and nearly constant latency, whereas alarm, video, and diagnostic traffic are bursty and primarily require BW. Placing every flow under the TAS mechanism is unnecessary, and for large networks, it is unrealistic. A more suitable architecture combines TAS with shapers such as CBS, ATS, or CQF, where TAS carries tightly constrained periodic streams and the other shapers serve traffic for which a fully TT schedule may be too expensive to configure~\cite{paul_iet_multi_shaper}. The difficulty with such combined mechanisms is that the shapers interact. Closing a TAS gate changes the service received by the CBS and ATS queues on the same egress port, and every queue or BW fraction assigned to one shaper becomes unavailable to the others. We are therefore interested in a joint configuration problem that decides and optimizes TTA, routing, queue assignment, GCL, and shaper parameters while considering the interdependencies~\cite{voica_multi_shaper_tsn}.

Next, consider a flow that crosses several domains. Each domain may find a feasible local schedule, while the complete path can still miss its end-to-end deadline. Time is spent at domain boundaries, synchronization errors must be accounted for, and the sum of the local latency budgets must remain within the end-to-end requirement. The domains may also differ in cycle lengths, shapers, and approaches to timing analysis. In addition, an administrative domain may be willing to disclose a delay bound and the capacity it offers, but not its topology or complete stream set. A centralized controller has more information to work with at the cost of a larger configuration problem and reduced domain autonomy. Hierarchical approaches instead exchange smaller abstractions; however, an abstraction may hide a feasible configuration~\cite{bhattacharjee_tnsm2024}. Based on these observations, we formulate our first finding.

\begin{figure}[!htbp]
\begin{tcolorbox}[colback=green!3,colframe=green!80!black,boxrule=0pt,leftrule=2.5pt,arc=1.5mm,left=4pt,right=4pt,top=4pt,bottom=4pt
]
    \textbf{Finding 1:} What is missing is a timing interface that is sufficiently detailed for end-to-end analysis, while not forcing a domain to reveal its internal configuration.
\end{tcolorbox}
\end{figure}

\begin{figure}[!t]
    \centering
    \includegraphics[width=\linewidth, trim={0cm 0cm 0cm 0cm}, clip]{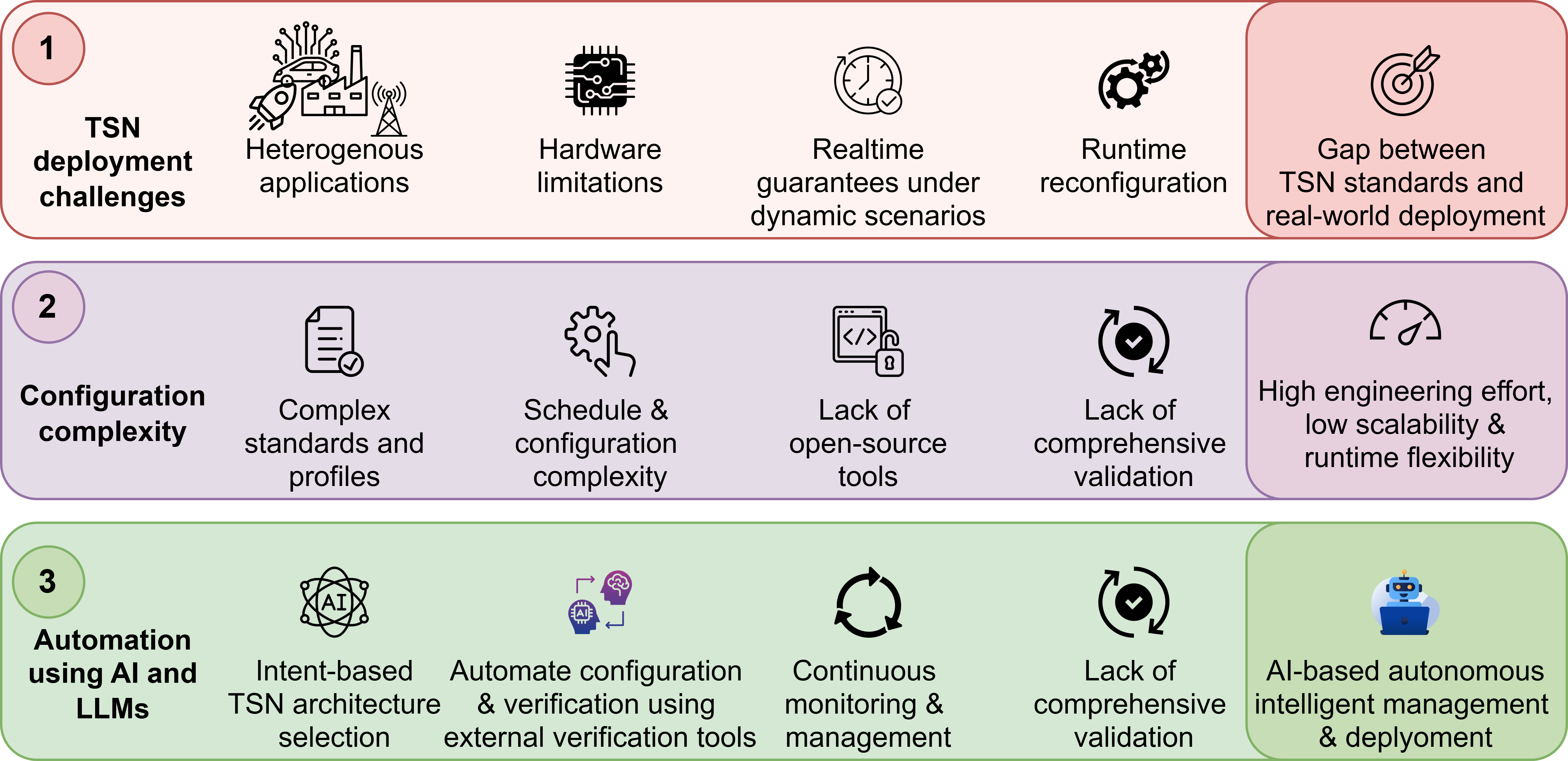}
    \caption{Roadmap and challenges towards end-to-end TSN deployment and management.}
    \label{fig:tsn_deployment}
\end{figure} 

A further limitation of TSN configuration (refer to Fig.~\ref{fig:tsn_deployment}) is the assumption that configuration occurs once, at design time. A deployed network does not remain fixed. Network devices are added, application modes are changed, links fail, and flows are added or removed. In such cases, waiting hours for a new global configuration is not an option. The admission or recovery algorithm must find a solution within a reasonable time and modify the existing configuration as little as possible. This is particularly challenging in multi-shaper networks, since moving a single TAS window may change the WCD of the ATS or CBS flows that share the egress port, and an update to an inter-domain flow may force several controllers to revise their local configurations. Incremental methods, however, reuse the previous solution and can admit changed flows much faster than a complete recomputation~\cite{gartner_comnet2023}, although the transition between the old and new configurations still needs to be analyzed~\cite{voica_multi_shaper_tsn}. This is discussed in further detail for the TAS mechanism in Section~\ref{sec:reconfiguration_by_marc_boyer}. 

On the other hand, TSN hardware imposes its own limitations. A solver can produce a mathematically valid schedule that the selected switches cannot store or execute, because egress ports have a limited number of queues and finite buffer memory. The GCL holds a bounded number of entries, and gate events can only be placed at the timing resolution supported by the device. When several shapers share a port, these limits become even more restrictive, since TAS, CBS, ATS, and CQF compete for the same queues and buffers, and shaper implementations differ in hardware cost~\cite{pruski_etfa2021}. Devices from different vendors may additionally support different parameter ranges or subsets of the standards. These hardware capabilities should be included as constraints in the problem formulation. Doing so may increase the size of the formulation, but omitting these constraints can result in undeployable configurations. These considerations motivate our second finding.

\begin{tcolorbox}[colback=green!3,colframe=green!70!black,boxrule=0pt,leftrule=2.5pt,arc=1.5mm,left=4pt,right=4pt,top=4pt,bottom=4pt]
    \textbf{Finding 2:} Hardware capabilities should be provided as inputs to configuration synthesis rather than being treated as checks applied after the synthesis step.
\end{tcolorbox}

Another challenge lies in the existing evaluation methodologies. The evaluation practice itself needs to evolve. Current evaluations report end-to-end delays and algorithm runtimes and stop there. These metrics provide limited insight into deployability in real-world systems. A schedule may satisfy all deadlines in the model and still not fit within the eight queues of an egress port or the GCL length of the respective switch, making the entire schedule unusable. The TSN community should additionally report the amount of device resources occupied by a configuration (queues, buffers, and GCL entries), the time required to admit or update a stream, and the number of devices and existing streams affected by such an update. Multi-domain studies should further specify their synchronization assumptions and the information exchanged between controllers. A configuration that is feasible in an abstract model is not necessarily one that can be installed, updated, and analyzed on the target network. The reported metrics should allow system engineers to understand this difference. Closing this abstraction gap requires shared benchmarks, constraints, best practices, and guidelines established through standards or organization-specific case studies. These artifacts can reduce assumptions and support consistent implementation. These aspects become particularly important as TSN profiles provide deployment guidance for specific application domains. However, several fundamental questions regarding mechanism selection, parameter configuration, and configuration validation remain open, motivating a closer examination of the current industrial TSN profiles and their limitations.

\section{Standardization Gaps Toward Deployable TSN}
\label{sec:configuration_and_standardization_efforts}
TSN profiles specify which features, options, defaults, and procedures apply within an application domain~\cite{60802}. Some profiles, like the automotive one, also include recommendations for configuring optional parameters~\cite{8021DG}. As outlined in Section~\ref{sec:deployment_and_runtime_challenges}, TSN profiles are intended to reduce the configuration space by providing concrete guidelines to the system engineer. Using the two recently published profiles, IEC/IEEE 60802~\cite{60802} and IEEE 802.1DG~\cite{8021DG}, as our reference, we highlight four practical questions: (i)~\textit{which mechanism or combination of mechanisms satisfies the application requirements}, (ii)~\textit{which parameter values should be assigned}, (iii)~\textit{how should the resulting configuration be verified}, and (iv)~\textit{whether the profile covers all use cases within its domain}.

The profiles provide the selection of features and options while deferring most concrete configurations and recommendations to system integrators, with synchronization in 60802 being a notable exception~\cite{60802,8021DG}. The profiles answer what a device vendor should implement rather than the system integrator's question about how to use it (see Table~\ref{tab:profile_gaps}).

\subsection{Mechanism Selection Gap}
IEC/IEEE 60802 defines eight traffic types, two conformance classes, and normative synchronization budgets of 1~$\mu$s and 100~$\mu$s over up to 100 hops for the working clock and the global time, respectively~\cite{60802}. The quantified traffic characteristics from the underlying use-case documents~\cite{ademaj2018,iic_traffic} are retained only as typical ranges or qualitative categories. Neither IEC/IEEE 60802 nor IEEE 802.1DG provides a comprehensive mandatory (normative) mapping from traffic types to concrete TSN mechanisms. IEEE 802.1DG binds mechanisms to traffic classes, but the class to which a given application stream belongs remains the integrator's decision. IEC/IEEE 60802 explicitly preserves this freedom, stating that ST is only one means of achieving time-triggered transmission and that its traffic-type categories permit different feature selections to reach the same goal~\cite{60802}. An Industry IoT Consortium (IIC) white paper maps traffic types to device capabilities but identifies itself as input to the standards, places configuration generally out of scope, and leaves even the choice between strict priority (SP) and ST for cyclic streams unspecified~\cite{iic_mapping}.

\begin{table}[!t]
\centering
\caption{Engineering decisions vs.\ what the two TSN profiles resolve (\harveyBallFull~resolved normatively, \harveyBallHalf~partially addressed or informative guidance only, \harveyBallNone~open or explicitly delegated).}
\label{tab:profile_gaps}
\begin{tabular}{@{}lcc@{}}
\toprule
\textbf{Engineering decision} & \textbf{60802} & \textbf{802.1DG} \\
\midrule
Traffic types and timing requirements & \harveyBallHalf & \harveyBallHalf \\
Synchronization accuracy and error budgets & \harveyBallFull & \harveyBallNone \\
Mandatory device feature set (conformance) & \harveyBallFull & \harveyBallFull \\
Mechanism selection per traffic type & \harveyBallHalf & \harveyBallHalf \\
Shaper and schedule parameter values & \harveyBallNone & \harveyBallHalf \\
Configuration model and synthesis toolchain & \harveyBallHalf & \harveyBallNone \\
Runtime reconfiguration & \harveyBallHalf & \harveyBallNone \\
End-to-end configuration verification & \harveyBallNone & \harveyBallNone \\
\bottomrule
\end{tabular}
\end{table}

IEC/IEEE 60802 adopts a revised variant of the taxonomy but not the mapping, offering a traffic-class mapping only as an example~\cite{60802}. In contrast, IEEE 802.1DG mandates ingress policing, together with CBS on the two highest traffic classes and ATS on at least one traffic class~\cite{8021DG}. TAS and FP are optional, while Frame Replication and Elimination for Reliability (FRER)~\cite{8021CB} is not used. Furthermore, time synchronization is only recommended and becomes mandatory only when the optional ST is supported, with five example synchronization mechanisms listed for the system integrator~\cite{8021DG}. These differences motivate Finding 3.

\begin{tcolorbox} [colback=green!3,colframe=green!70!black,boxrule=0pt,leftrule=2.5pt,arc=1.5mm,left=4pt,right=4pt,top=4pt,bottom=4pt]
    \textbf{Finding 3:} Current TSN profiles do not provide a definitive mapping between application requirements, traffic types, and appropriate TSN mechanisms. Consequently, mechanism selection remains largely dependent on the expertise of system integrators.
\end{tcolorbox}

We argue that the selection gap persists because TSN mechanisms overlap in functionality and experts optimize for different objectives, such as tight latency and jitter guarantees, configuration simplicity, low adoption risk, or hardware cost, with no single mechanism being optimal for all objectives. Disagreement is therefore not only a matter of incomplete understanding, as standardizing on a mechanism that best serves one objective can make other objectives harder to achieve. For the highest-priority class in the absence of overload, CBS and ATS shaping do not improve the worst-case latency but instead smooth the shaped class's bursts and improve the average latency of lower-priority classes~\cite{turner_aec2023}. Neither profile distinguishes between worst-case guarantees and burst regulation, leaving system integrators to infer what each mandated mechanism actually contributes. Moreover, configuring CBS according to the standard's own idle-slope rule can cause worst-case analysis to fail for most configurations that are in fact schedulable~\cite{ashjaei2017}. ATS, meanwhile, provides only limited, load-dependent advantages over CBS~\cite{luxi_tnsm}. Together, these observations further complicate the selection of the appropriate shaping mechanism or combination of mechanisms.

Experimental results also reveal significant differences among mechanisms. In one study on automotive 100BASE-T1 hardware, a low-priority interferer increases the measured two-hop worst-case latency from approximately 40--50~$\mu$s to 289~$\mu$s under SP and to 269~$\mu$s under CBS. Under TAS, however, the latency stays at 47~$\mu$s~\cite{janker_measurements}. Earlier work likewise identified the shaping mechanism for control traffic as an open question, warning that an unstructured collection of mechanisms could drive adopters back to FlexRay or CAN FD~\cite{jochim2013}.

IEC/IEEE 60802 requires TAS at 100~Mbit/s and 1~Gbit/s in its feature-rich conformance class. It also requires FP, unconditionally for bridges and conditionally for end stations, but does not reference ATS~\cite{60802}. Domain differences alone do not fully explain the divergence between the mechanism choices of the two profiles. Within a single year, contributions to the automotive group recommended CBS with TAS and preemption as last resorts~\cite{pannell_choosing}, preemption before TAS due to risk considerations~\cite{budweiser_riskaverse}, and a hardened audio-video bridging (AVB) subset containing neither~\cite{budweiser_avbprofile}. Some training material even argues that TAS and CQF have no role in a converged in-vehicle backbone network~\cite{turner_aec2023}. Formal bounds exist for all mechanisms, but TAS additionally enables joint task/network scheduling to minimize alignment delays and tightly control jitter~\cite{craciunas_rts16}.

Whether a profile covers all target-domain use cases remains a practical question. Between 2019 and 2022, at least three incompatible traffic-type taxonomies were proposed~\cite{chen_traffictypes,turner_trafficclass,gopal_trafficclass}, before the published IEEE 802.1DG profile settled on an informative characterization using order-of-magnitude figures~\cite{8021DG}.

The asynchronous foundation of IEEE 802.1DG assumes independently developed domains that exchange rate contracts without a common clock~\cite{specht_ubs_auto}. Software-defined vehicles centralize integration, and automated driving in particular may require a common notion of time for sensor fusion~\cite{lobello2019, tsn_automotive}. Moreover, the cause-effect chains of ADAS/AD functions accumulate sampling delays at every task and network boundary~\cite{becker2017, duerr2019}, and their host systems are highly utilized~\cite{CraciunasRTSOPS26}. Asynchronous mechanisms can bound such chains, but minimizing them requires co-scheduling the task and network domains~\cite{craciunas_rts16}, a capability that only TAS provides. The profiles have thus not resolved the selection problem but have only institutionalized it.

\subsection{Parameterization Gap}
Even after the mechanism set is fixed, deriving suitable parameters remains a difficult task for the system integrator. IEEE 802.1DG provides closed-form policing parameters, with an idle slope equal to the sum of the stream rates~\cite{8021DG}. However, this corresponds to the same allocation under which the worst-case analysis fails~\cite{ashjaei2017}. Beyond this suggested policing rule, neither profile specifies how scheduler and shaper parameters should be derived from application requirements. Such derivation methods are valid design choices, and we do not argue that a profile should mandate a specific method. Instead, what is missing is a higher-level framework that provides a standardized way to express the requirements that a configuration must satisfy and a means to check the result independently of the tool that produced it. IEC/IEEE 60802 demonstrates that this separation is feasible by standardizing device capabilities in the form of a digital data sheet while leaving open the computation that uses these capabilities~\cite{60802}. These considerations motivate Finding 4.

\begin{tcolorbox}[
colback=green!5,
colframe=green!80!black,
boxrule=0pt,
leftrule=2.5pt,
arc=1.5mm,
left=4pt,
right=4pt,
top=4pt,
bottom=4pt
]
    \textbf{Finding 4:} The selection of a TSN mechanism alone does not determine network guarantees. The resulting performance depends on parameter configuration, traffic characteristics, and interactions between mechanisms. 
\end{tcolorbox}

For TAS, GCL synthesis is NP-hard~\cite{CraciunasRTNS16}, and exact formulations reach their practical limits on moderately sized networks~\cite{ilp_limitations}. Nevertheless, tailored heuristics can schedule instances with 2000 nodes and more than 10\,000 flows within one hour~\cite{vlk_cor2022}, and systematic experimental evaluations have studied the schedulability and scalability of such approaches~\cite{xue_rtas24}. An in-vehicle backbone consisting of a few switches is therefore orders of magnitude smaller than these evaluated instances. Design-time guarantees for CBS or ATS also require assigning streams to at most $8$ traffic classes, which is itself a discrete optimization problem~\cite{voica_traffic_assignment}, and deriving the coupled shaper parameters. Together with subsequent worst-case analysis, for which computing tight bounds can be NP-hard~\cite{bouillard_tight}, the overall problem remains computationally complex. Selecting an asynchronous mechanism therefore shifts complexity from GCL synthesis to verification. Where no analysis is mandated, this complexity is effectively dropped rather than solved. This tradeoff motivates Finding 5.

\begin{tcolorbox}[
colback=green!3,
colframe=green!70!black,
boxrule=0pt,
leftrule=2.5pt,
arc=1.5mm,
left=4pt,
right=4pt,
top=4pt,
bottom=4pt
]
    \textbf{Finding 5:} Choosing simpler mechanisms does not remove complexity. It only shifts complexity from schedule synthesis to worst-case analysis and verification.
\end{tcolorbox}

The complexity warnings in IEEE 802.1DG concern only features that are optional or not used, namely ST and the stream identification used for frame replication~\cite{8021DG}, while IEC/IEEE 60802 mandates TAS without discussing its complexity~\cite{60802}. Moreover, the complexity argument against TAS in~\cite{turner_aec2023} is based on work that also proposes heuristics to make the synthesis tractable (see~\cite{pop_iet2016}). Thus, the complexity that influences mechanism selection is documented by research but is not weighed symmetrically by the two profiles. Both profiles define certain conformance floors, such as minimum queue and stream-filter counts, but neither guarantees a minimum capacity for GCLs. IEC/IEEE 60802 limits gate granularity but treats list lengths and cycle times as self-declared datasheet capabilities~\cite{60802}. Similarly, the entry counts in IEEE 802.1DG are provided only as guidance and, as the standard states, do not constitute conformance requirements~\cite{8021DG}. These challenges motivate Finding 6.

\begin{tcolorbox}[
colback=green!3,
colframe=green!70!black,
boxrule=0pt,
leftrule=2.5pt,
arc=1.5mm,
left=4pt,
right=4pt,
top=4pt,
bottom=4pt
]
    \textbf{Finding 6:} TSN profiles leave parameter values, optimization, and deployment decisions largely implementation-dependent.
\end{tcolorbox}

The question of configuration responsibility is also only partially addressed. IEC/IEEE 60802 adopts the fully centralized model of 802.1Qcc~\cite{8021Qcc}, requiring NETCONF/YANG management. Resource-constrained devices without native NETCONF support can only be integrated through proxies using user-specific protocols~\cite{60802}. However, the computation of schedules and paths performed by the Centralized Network Configuration (CNC) entity remains implementation-dependent. Its interface to engineering tools is user-specific~\cite{60802}, while IEEE 802.1DG specifies no configuration protocol at all~\cite{8021DG}. Consequently, the actual computation of the network configuration remains outside the scope of the standards. Although IEC/IEEE 60802 mandates candidate-based configuration updates with rollback and isolates newly added stations until the CNC admits their traffic, the method used to compute incremental updates remains unspecified.

\subsection{Configuration Verification Gap}
Conformance testing certifies devices, not network configurations. The common test specification currently being developed for IEC/IEEE 60802 defines its testing scope at the level of stations and their components~\cite{tiacc2025}, and the certification programs built on that baseline inherit the same scope. They do not provide a standardized method to verify whether a particular assignment of streams, parameters, and schedules satisfies application-level requirements. This gap motivates Finding 7.

\begin{figure}[!htbp]
\begin{tcolorbox}[
colback=green!3,
colframe=green!70!black,
boxrule=0pt,
leftrule=2.5pt,
arc=1.5mm,
left=4pt,
right=4pt,
top=4pt,
bottom=4pt
]
    \textbf{Finding 7:} Device conformance has a clear owner, i.e., the vendor that tests a device against a specification. A deployed configuration spans multiple mechanisms, vendors, and integrators, leaving no single party responsible for its verification and no current program covering it.
\end{tcolorbox}
\end{figure}

The analytical guidance provided by current standards is also incomplete. Even the bound provided by the AVB profile does not cover its worst-case traffic pattern~\cite{turner_aec2023,8021BA}. The CBS analysis in 802.1Q Annex L excludes any interruption to BW availability caused by gate operation~\cite{8021Q}, and IEEE 802.1DG declines to give guidance on how such combinations affect network behaviour~\cite{8021DG}. These limitations motivate Finding 8.

\begin{figure}[!htbp]
\begin{tcolorbox}[
colback=green!3,
colframe=green!70!black,
boxrule=0pt,
leftrule=2.5pt,
arc=1.5mm,
left=4pt,
right=4pt,
top=4pt,
bottom=4pt
]
    \textbf{Finding 8:} No standardized instrument validates whether a complete configuration satisfies the intended guarantees.
\end{tcolorbox}
\end{figure}

The selection and parameterization gaps share a common root, which may also explain the expert disagreement. Guarantees depend on parameter values rather than on the mechanism alone. Therefore, the configuration space grows exponentially with the number of parameters rather than with the number of permitted mechanisms. These observations reveal three needs, none of which requires a profile to mandate a mechanism:
\begin{enumerate}[leftmargin=*]
    \item \textbf{Methodology is required:} Validated configuration patterns that map traffic classes to mechanism combinations, together with the required analyses and hardware assumptions. The closest existing attempts~\cite{iic_mapping,chahed2023} are not normative documents. However, the ongoing IEEE project on shaper parameter settings~\cite{8021Qdq} demonstrates that this need has been recognized.
    \item \textbf{Tooling is required:} Open configuration synthesis approaches that consume the hardware model already standardized by IEC/IEEE 60802 and validate configurations rather than only devices. These requirements connect to Section~\ref{sec:deployment_and_runtime_challenges}, Section~\ref{sec:reconfiguration_by_marc_boyer}, and Section~\ref{sec:llm_tsn_deployment}.
    \item \textbf{Transparent guidance is required:} Published negative results identifying which mechanisms are substitutes, which combinations do not compose, and which mechanisms are unsuitable for specific use cases.
\end{enumerate} 

\section{Toward a Deployment-Oriented Solution}
\label{sec:reconfiguration_by_marc_boyer}
Configuration does not end with the initial deployment. Even after selecting the optimal mechanisms and generating the initial configuration, there are several reasons why a TSN network must be reconfigured during operation. Streams may be added or removed, an application may switch its operating mode, a device may be replaced during maintenance, and, in most projects, the design itself is extended incrementally. In addition to these planned changes, the network must recover from link, port, or device faults at runtime. Such scenarios impose stringent requirements: computing and installing a new configuration must be completed within a bounded time, while traffic that does not rely on the failed component must continue to satisfy its timing and reliability requirements. One approach to reducing both computation and deployment effort is incremental reconfiguration, i.e., modifying as few streams and devices as possible. However, minimizing the number of affected elements alone is insufficient, as timing and loss guarantees must also be maintained throughout the transition from the old configuration to the new one~\cite{boyer_henia_ecrts2025,voica_multi_shaper_tsn}.

\subsection{Closed Loop Runtime Reconfiguration}
It is useful to view runtime reconfiguration as a loop consisting of four steps: \emph{detect--compute--deploy--verify}. The detection step may appear straightforward, but it is not, as the system must determine whether a component has failed or a stream has changed its requirements, and identify which flows are affected before any further action can begin. Computation is the aspect most familiar to the scheduling community, i.e., selecting the admitted flows together with their routes, queues, shapers, GCLs, and parameters; however, the difference is that the algorithm now operates under the processing and memory constraints of the deployed system. The resulting configuration must then be deployed to the bridges and activated. In the final step, the controller verifies that each device has accepted its configuration, that the installed state is consistent across devices, and that the timing constraints continue to be satisfied. A fast scheduling algorithm provides limited benefit if fault detection is slow or if the update protocol can leave the network in an inconsistent state~\cite{boyer_henia_ecrts2025,pang_tii2021}. These observations motivate Finding 9.

\begin{tcolorbox}[
colback=green!3,
colframe=green!70!black,
boxrule=0pt,
leftrule=2.5pt,
arc=1.5mm,
left=4pt,
right=4pt,
top=4pt,
bottom=4pt
]
    \textbf{Finding 9:} A fast scheduling algorithm alone is insufficient for runtime TSN reconfiguration. Practical solutions require an integrated \emph{detect--compute--deploy--verify} workflow to ensure consistent updates and continued satisfaction of timing requirements.
\end{tcolorbox}

\subsection{Failure Recovery and Degraded Modes}
Where should fault tolerance originate? The standards provide precomputed redundancy mechanisms. FRER (IEEE 802.1CB~\cite{8021CB}) duplicates frames over disjoint paths, and backup configurations can be prepared for anticipated faults. Recovery is then fast; however, the BW, memory, and configuration resources required for the backups are consumed regardless of whether a fault occurs. The alternative is to reroute and reschedule online after the failure has been detected; resources are then consumed only when needed, at the cost of bounding the computation and deployment times. Neither approach is sufficient on its own. The number of backups grows combinatorially with network size and the number of faults to be tolerated, while after a severe fault, the remaining network may no longer provide the nominal level of service. Kong et al.~\cite{kong_access2021} study runtime recovery of TT traffic under such conditions. When available resources are insufficient, the recovery algorithm either maintains full functionality, moves selected applications into degraded mode, or rejects lower-utility traffic. This utility should be modeled separately from the 802.1Q priority, which also influences schedulability~\cite{boyer_henia_ecrts2025}. These tradeoffs motivate Finding 10.

\begin{tcolorbox}[
colback=green!3,
colframe=green!70!black,
boxrule=0pt,
leftrule=2.5pt,
arc=1.5mm,
left=4pt,
right=4pt,
top=4pt,
bottom=4pt
]
\textbf{Finding 10:} Precomputed redundancy and online recovery alone are insufficient for scalable TSN fault tolerance. Deployable TSN networks require resource-aware fault tolerance strategies that dynamically balance reliability, resource availability, and application criticality.
\end{tcolorbox}

\subsection{Consistent Updates}
A feasible new configuration does not define the transition toward that configuration. Bridges activate routing tables and GCLs at different times, while frames from the previous configuration may still be queued. Such frames may encounter stale forwarding states and be dropped at an intermediate bridge, or may occupy a TAS window assigned to another stream, delaying subsequent frames and violating the schedule. An update method must hence define the order and activation times of device changes while accounting for in-flight and queued frames. The necessary guarantees must cover the transition interval as well as the two steady-state configurations. Conflict-free update methods address this problem for scheduled flows~\cite{pang_tii2021}. However, when routing, multiple shapers, and device-specific configuration interfaces change simultaneously, consistent deployment remains a challenge~\cite{boyer_henia_ecrts2025}. These requirements motivate Finding 11.

\begin{figure}[!htbp]
\begin{tcolorbox}[
colback=green!3,
colframe=green!70!black,
boxrule=0pt,
leftrule=2.5pt,
arc=1.5mm,
left=4pt,
right=4pt,
top=4pt,
bottom=4pt
]
\textbf{Finding 11:} Consistent TSN deployment requires update mechanisms that coordinate configuration changes across devices and preserve timing guarantees throughout the transition between configurations.
\end{tcolorbox}
\vspace{-0.3cm}
\end{figure}

\subsection{Design Point Solution}
\begin{figure}[!t]
  \centering
  \begin{subfigure}{\linewidth}
    \centering
    \includegraphics[width=0.98\linewidth,trim={0 0 0 0},clip]{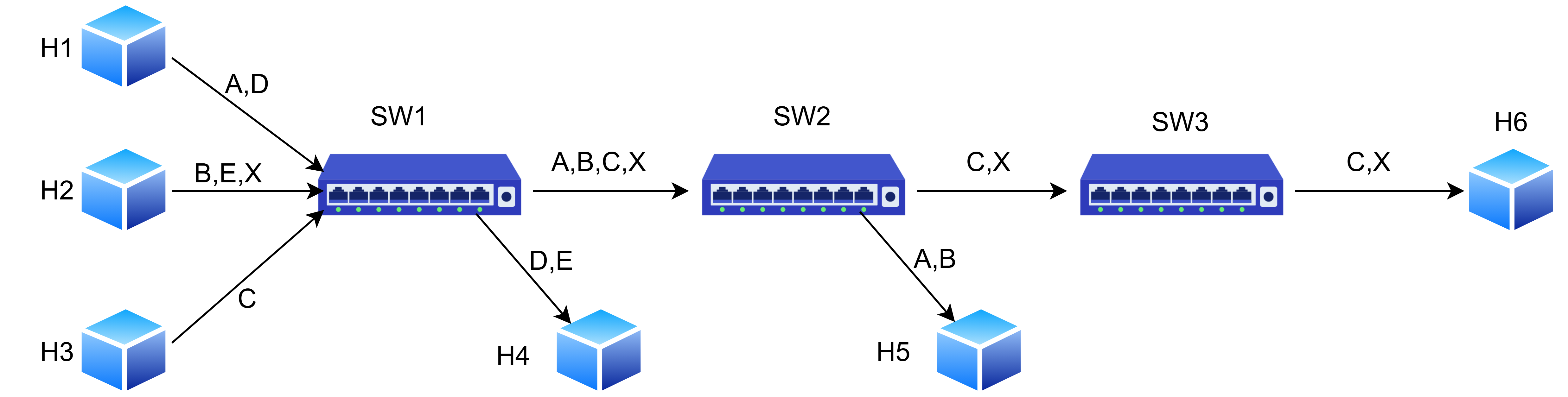}
    \caption{Topology; the edge labels list the routed streams, including the new stream~$X$.}
    \label{fig:wpex-summary-topo}
    \vspace*{3pt}
  \end{subfigure}
  \begin{subfigure}{\linewidth}
    \centering
    \includegraphics[width=0.98\linewidth, trim={0 0 0 0},clip]{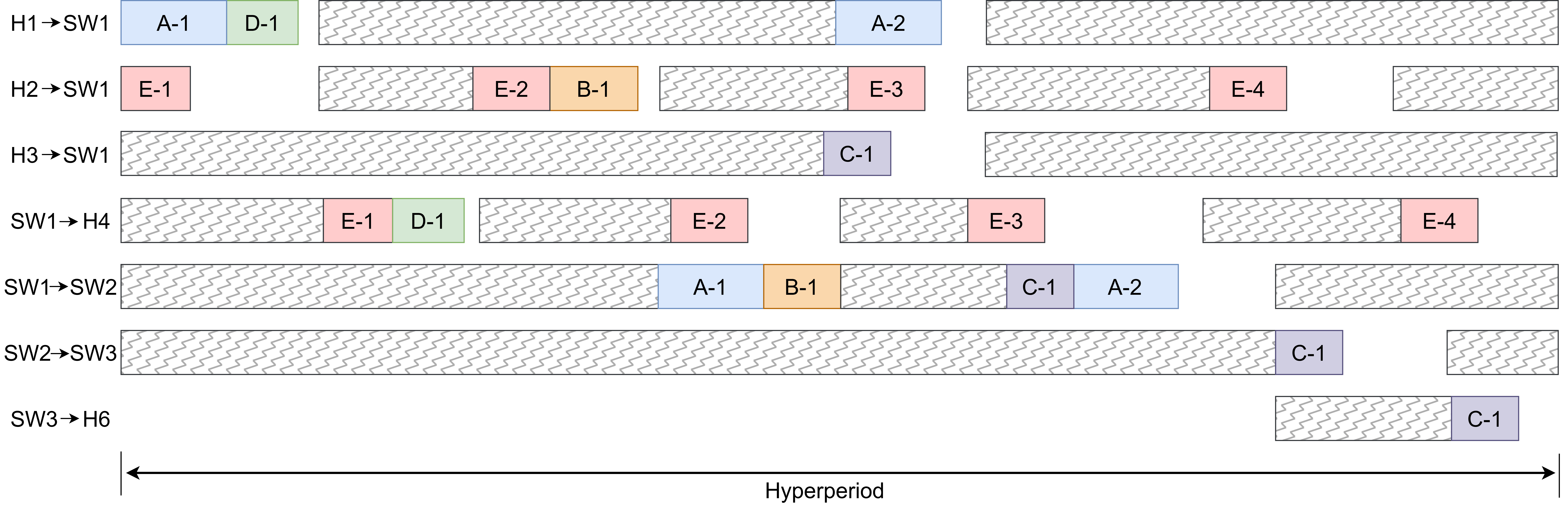}
    \caption{Windows enlarged to provide slack.}
    \label{fig:wpex-summary-slack}
    \vspace*{3pt}
  \end{subfigure}
  \begin{subfigure}{\linewidth}
    \centering
    \includegraphics[width=0.98\linewidth,trim={0 0 0 0},clip]{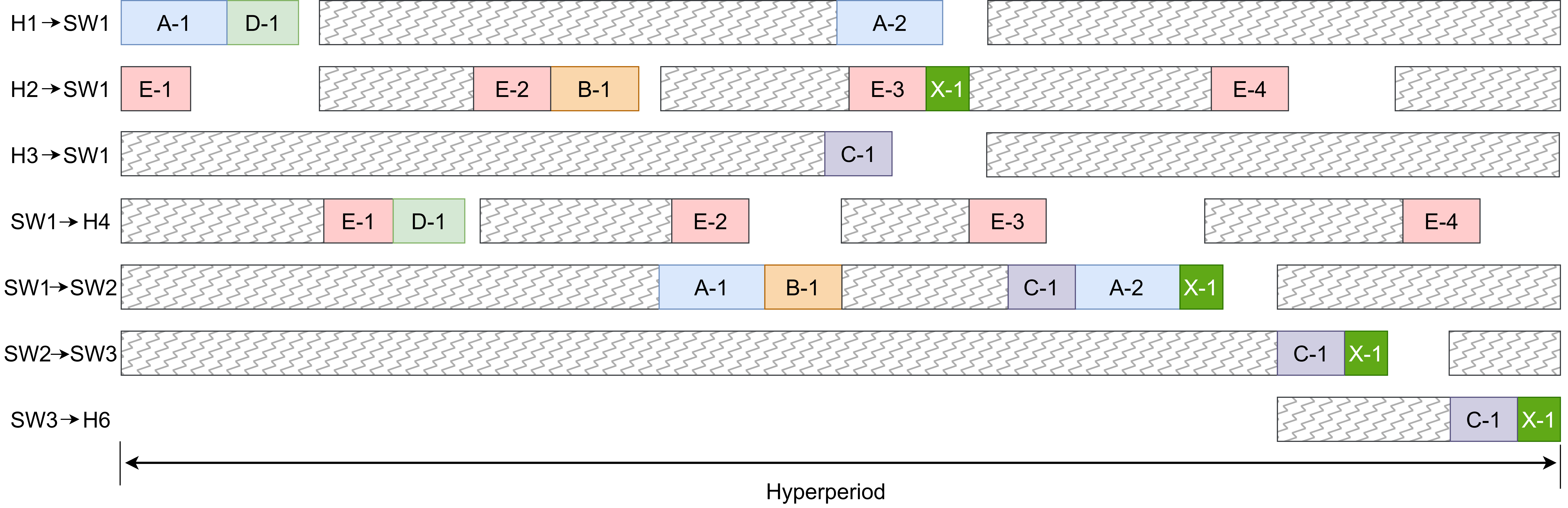}
    \caption{Stream~$X$ admitted using the available windows.}
    \label{fig:wpex-summary-update}
  \end{subfigure}
  \caption{WPEx example: (a) the topology; (b) enlarged TAS windows provide slack; (c) a new stream~$X$ uses the available windows without changing the GCL.}
  \label{fig:wpex-summary}
\end{figure}

Reconfiguration capabilities can also be incorporated into the initial schedule. The Windows Precedence Exclusion (WPEx) method introduces this concept for TAS~\cite{TAS-Reconf-WPEx-RTNS-2025}. WPEx assigns sets of frames to precedence-separated windows and reserves slack by spreading and enlarging the windows. Admitting or rerouting a stream then becomes a matter of finding a path whose sequence of windows still has available capacity. The GCL remains unchanged, only the routing and stream-to-queue state are updated. Fewer device configuration elements are modified, reducing the exposure to inconsistent GCL activation. A WPEx schedule also tolerates missing frames, which is relevant when a stream is removed or interrupted by a fault. Under protective gating, lower-priority traffic may use the unused portion of an enlarged window left by TAS frames, subject to the assumed priority and arrival conditions~\cite{Protective-Gating}. \Cref{fig:wpex-summary} illustrates the topology and the two steps: the enlarged windows with their reserved slack and the admission of a new stream~$X$ into that slack (shown in green).

WPEx represents one possible design approach in this direction, addressing a specific problem under specific assumptions. A WPEx-compatible TAS schedule must exist as the starting point, and admission succeeds only if the selected path still has sufficient slack. This approach assumes a centralized configuration entity, leaving fault detection and fault tolerance of the controller out of scope, and relies on consistent updates of routing tables even though the GCL remains unchanged. In this approach, we have not evaluated whether it extends to ports where multiple shapers interact, to paths that cross administrative domains, or to devices with heterogeneous hardware capabilities. Together with flexibility-based admission~\cite{TAS-Flexibility}, WPEx demonstrates that an initial schedule can reserve resources and structure that reduce the cost of future changes. Runtime TSN reconfiguration is therefore a control loop whose phases require bounded execution times and validation mechanisms, rather than merely a scheduling problem to be solved more quickly. Several challenges remain open, as summarized in Finding 12, some of which may be addressed through ML-based automation. 

\begin{tcolorbox}[
colback=green!3,
colframe=green!70!black,
boxrule=0pt,
leftrule=2.5pt,
arc=1.5mm,
left=4pt,
right=4pt,
top=4pt,
bottom=4pt
]
    \textbf{Finding 12:} Incremental timing analyses that account for interacting shapers are largely absent, failure-aware coordination among domain controllers remains unresolved, and update protocols that preserve guarantees during deployment are still in their early stages.
\end{tcolorbox}

Automated management has a natural role in addressing these challenges, provided that the roles and responsibilities remain clearly separated. An automated agent may propose recovery actions, coordinate the reconfiguration workflow, and drive the configuration tools, while deterministic timing analysis and device-level validation determine whether a proposed action can be activated. The large language model (LLM)-assisted architecture presented in the next section is built around precisely this separation, combining AI-assisted tools with deterministic verification to preserve correctness and timing guarantees.

\section{AI-Aided TSN Deployment}
\label{sec:llm_tsn_deployment}
The configuration and management challenges discussed in the previous sections motivate the exploration of learning-based approaches for TSN deployment. As network size and configuration complexity increase, exploring the configuration space requires evaluating a large number of candidate solutions while accounting for routing, scheduling, traffic assignment, and hardware constraints. Formal verification techniques remain essential for providing deterministic guarantees, but repeatedly applying them during reconfiguration for search space exploration can become a major computational bottleneck~\cite{reusch2023}. Graph Neural Networks (GNNs) are particularly promising in this context because they can represent network topology~\cite{drl_gnn_tsn}. Recent advances in deep reinforcement learning (DRL) have demonstrated their potential for TSN routing and scheduling~\cite{drl_jrsp_mixed_flows_hao, deepcqf_icc_cheng, deepscheduler}, and TTA~\cite{rubi_icc} in complex network scenarios. Moreover, generative AI and LLMs have emerged as a promising tool to support system engineers with configuration and debugging. \cite{survey_llm_network_operation} surveys the advantages and roles of LLMs in intelligent network management. Prior work has explored using LLMs to extract information from complex network documents~\cite{prosper_llm}, and has proposed a verified programming method for LLMs to generate router configurations~\cite{llm_configure_router}. TSN management and configuration are largely performed manually by domain experts and system engineers, requiring significant time and effort and carrying a high risk of errors from human judgment failures. LLMs can improve efficiency here by detecting errors and monitoring network logs in real time~\cite{survey_llm_network_operation}, and by learning TTA, traffic patterns, TSN architectures, and other design information to dynamically suggest changes to system engineers based on network conditions~\cite{survey_llm_network_operation}. These opportunities motivate Finding 13.

\begin{figure}[!t]
    \centering
    \includegraphics[width=\linewidth, trim={0.5cm 0.7cm 0.5cm 0.5cm}, clip]{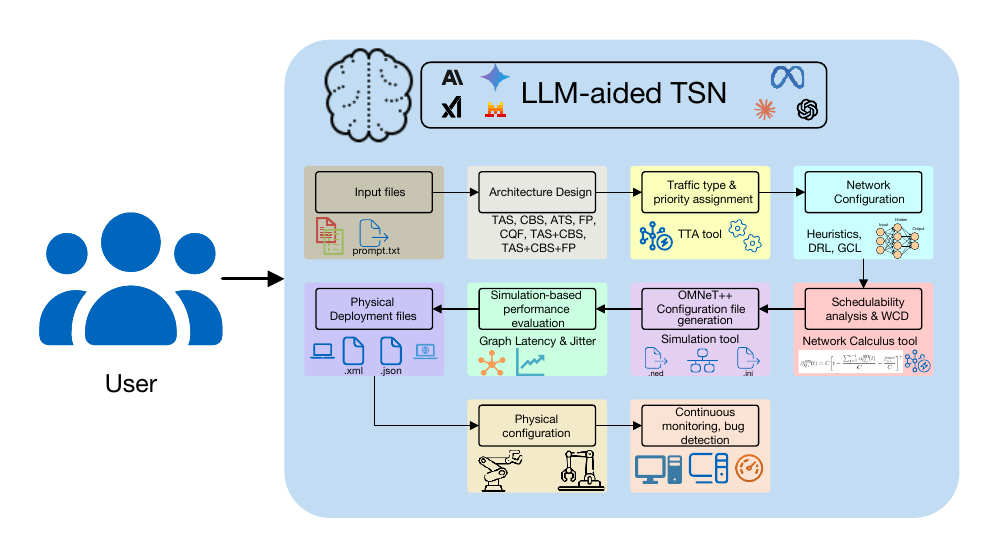}
    \caption{A vision and a conceptual framework illustrating LLM-assisted TSN management and deployment.}
    \label{fig:tsn_orchestrator}
    \vspace{-0.3cm}
\end{figure}

\begin{tcolorbox}[
colback=green!3,
colframe=green!70!black,
boxrule=0pt,
leftrule=2.5pt,
arc=1.5mm,
left=4pt,
right=4pt,
top=4pt,
bottom=4pt
]
    \textbf{Finding 13:} LLMs can meaningfully support multiple stages of the end-to-end TSN deployment lifecycle (refer to Fig.~\ref{fig:tsn_orchestrator}), from design to validation and troubleshooting. 
\end{tcolorbox}

However, using LLMs for network configuration is not straightforward and presents additional challenges~\cite{netconfeval}. LLMs cannot magically replace system engineers by fully automating management and configuration. Rather, they can meaningfully reduce the overall time spent on TSN deployment. The emerging capabilities of LLMs present an opportunity to explore their applicability across different stages of TSN management and orchestration and to identify the use cases in which they can provide meaningful benefits. As a first step, \cite{rubi_vtc_2026} proposed the first end-to-end TSN management framework. Building on this, \cite{rubi_tsnbench} presented TSNBench, benchmarking 16 state-of-the-art LLMs on multiple-choice and open-ended TSN questions. Verification and approval of LLM-provided solutions remain at the discretion of domain experts. Each configuration or design choice is accompanied by a model-generated confidence score, helping the system engineer decide whether to accept the model's output.

To fully realize the potential of LLMs for TSN management and deployment, our community must advance several open research directions. As demonstrated in \cite{rubi_tsnbench}, off-the-shelf LLMs fail to assist with schedulability analysis and open-ended questions. The models, however, exhibit a good understanding and capability in solving multiple-choice questions related to TSN. This gap highlights clear directions for future research aimed at optimizing and improving LLMs for TSN applications. First, models should be trained on TSN documents, such as the IEEE 802.1 standards, to strengthen their domain knowledge. To further facilitate the configuration of TSN hardware and switches, company-specific documentation can be incorporated through retrieval-augmented generation (RAG) or, where appropriate, model fine-tuning, enabling network engineers to interact with LLMs and receive configuration guidance for new TSN hardware deployments. This approach would ultimately reduce manual effort and operational overhead. However, LLM outputs cannot be used as-is without appropriate verification. Therefore, we must develop new tools or leverage existing ones to verify LLM outputs before deploying them in real-world scenarios. Existing mathematical models~\cite{ilp_limitations, silviu_gate_array, CraciunasRTNS16}, heuristic and metaheuristic-based algorithms~\cite{mcqf_paul, voica_traffic_assignment, rubi_iotj}, NC tools~\cite{luxi_tnsm, rubi_rtss, luxi_avb_rtas, luxi_multiple_avb}, and ML-based solutions~\cite{deepcqf_icc_cheng, deepscheduler, rubi_icc} provide strong foundations for this verification. LLMs can generate code for NC-based evaluation, constraints for scheduling and configuration algorithms, code for generating the final output, and network simulation files. However, providing the necessary guarantees for LLM outputs requires additional tools to verify them. Automated verification is therefore crucial for utilizing LLMs effectively in TSN management and orchestration frameworks. These verification requirements motivate Finding 14.

\begin{figure}[!htbp]
\begin{tcolorbox}[
colback=green!3,
colframe=green!70!black,
boxrule=0pt,
leftrule=2.5pt,
arc=1.5mm,
left=4pt,
right=4pt,
top=4pt,
bottom=4pt
]
    \textbf{Finding 14:} LLM-generated configurations cannot be directly deployed for TSN management without external verification from traditional tools.
\end{tcolorbox}
\end{figure}

\section{Research Roadmap Toward Deployable TSN}
\label{sec:research_directions}
To make TSN deployable, there are many open research directions. Here, we summarize the identified research directions for the short, medium, and long term.

\noindent \textbf{Short-term:} In the short term, research should focus on developing AI-assisted scheduling and configuration algorithms that account for hardware limitations while generating solutions. Lifecycle-aware scheduling and configuration represent the first critical step toward making TSN deployable. 

\noindent \textbf{Medium-term:} Over the next four to five years, research should focus on developing automated TSN deployment pipeline architectures. AI-assisted verification and configuration tools that take natural language as input and generate configuration and scheduling tables, verified by mathematical models, represent the future of TSN network management. Such tools will assist network engineers with deployment, reducing the burden on domain experts. Runtime reconfiguration is not a new concept, yet a fully deployable, runtime-reconfigurable solution for all stages of the TSN lifecycle remains out of reach. 

\noindent \textbf{Long-term:} The long-term goal is to develop an end-to-end AI-assisted TSN deployment pipeline that provides verifiable solutions and configuration artifacts, uses mathematical models to verify generated solutions, automatically monitors the deployed network, and assists with runtime reconfiguration. Developing agentic-TSN, where network devices can self-configure and self-heal to support a plug-and-play architecture, is essential for modern communication infrastructure. Moreover, the standardization community must work toward providing standardized datasets, benchmarks, and, most importantly, guidelines for the unified development of configuration, scheduling, and optimization solutions.

\section{Conclusion}
\label{sec:conclusion}
Deploying TSN is an end-to-end engineering problem that extends well beyond the design of individual mechanisms. This paper examined the TSN deployment lifecycle and identified where current practice falls short. Existing tools ignore shaper interactions and hardware constraints. Industrial profiles select mechanisms without resolving their configuration. Runtime reconfiguration lacks bounded and consistent update paths. AI-assisted management remains trustworthy only when gated by deterministic verification. Closing these gaps requires treating configuration, evaluation, and automation as stages of a single workflow rather than isolated research problems. Realizing such a workflow demands community infrastructure that includes open tools, reference benchmarks, and transparent guidance on which mechanism belongs in which use case. The long-term goal is agentic-TSN networks that self-configure and self-heal while remaining verifiable through mathematical analysis. In future work, we will develop the framework and requirements for agentic-TSN toward adaptive, runtime-reconfigurable TSN deployment.

\bibliographystyle{IEEEtran}
\bibliography{reference}
\end{document}